\documentclass[12pt,preprint]{aastex}

\slugcomment{ApJ accepted.}

\shorttitle{High-Contrast NIR Imaging of MWC 480}
\shortauthors{Kusakabe et al.}

\begin{document}


\title{High-Contrast NIR Polarization Imaging of MWC480}


\author{N. Kusakabe\altaffilmark{1}, C. A. Grady\altaffilmark{2}
M. L. Sitko\altaffilmark{3,4,5}, J. Hashimoto\altaffilmark{1}, T. Kudo\altaffilmark{1}, M. Fukagawa\altaffilmark{10},}
 \author{T. Muto\altaffilmark{12,13}, J. P. Wisniewski\altaffilmark{11}, M. Min\altaffilmark{14}, S. Mayama\altaffilmark{32}, C. Werren\altaffilmark{3,4}, A. N. Day\altaffilmark{3,4,6}, L. C. Beerman\altaffilmark{3,4,7}, D. K. Lynch\altaffilmark{4,8}, R. W. Russell\altaffilmark{4,8}, S. M. Brafford\altaffilmark{4,9}, M. Kuzuhara\altaffilmark{30}, T. D. Brandt\altaffilmark{21}, L. Abe\altaffilmark{15}, 
W. Brandner\altaffilmark{16}, J. Carson\altaffilmark{17}, S. Egner\altaffilmark{18}, M. Feldt\altaffilmark{16}, M. Goto\altaffilmark{31}, O. Guyon\altaffilmark{18}, Y. Hayano\altaffilmark{18}, M. Hayashi\altaffilmark{19}, S. S. Hayashi\altaffilmark{18},T. Henning\altaffilmark{16}, K. W. Hodapp\altaffilmark{20}, M. Ishii\altaffilmark{19}, M. Iye\altaffilmark{1}, M. Janson\altaffilmark{21}, R. Kandori\altaffilmark{1}, G. R. Knapp\altaffilmark{21}, T. Matsuo\altaffilmark{22}, M. W. McElwain\altaffilmark{23}, S. Miyama\altaffilmark{1}, J.-I. Morino\altaffilmark{1}, A. Moro-Martin\altaffilmark{21,24}, T. Nishimura\altaffilmark{18}, T. -S. Pyo\altaffilmark{18}, H. Suto\altaffilmark{1}, R. Suzuki\altaffilmark{1}, M. Takami\altaffilmark{25}, N. Takato\altaffilmark{18}, H. Terada\altaffilmark{18}, C. Thalmann\altaffilmark{26}, D. Tomono\altaffilmark{18}, E. L. Turner\altaffilmark{21,27}, M. Watanabe\altaffilmark{28}, T. Yamada\altaffilmark{29}, 
H. Takami\altaffilmark{18}, T. Usuda\altaffilmark{18}, and M. Tamura\altaffilmark{1}
}


\altaffiltext{1}{National Astronomical Observatory, 2-21-1 Osawa, Mitaka, Tokyo 181-8588}
  \email{nb.kusakabe@nao.ac.jp}
\altaffiltext {2} {Eureka Scientific, 2452 Delmer St., Suite 100, Oakland CA 96402}
\altaffiltext{3} {Department of Physics, University of Cincinnati, Cincinnati OH 45221, USA}
\altaffiltext{4} {Guest Observer, Infrared Telescope Facility} 
\altaffiltext{5} {Space Science Institute, 4750 Walnut St., Suite 205, Boulder CO 80301, USA}
\altaffiltext{6} {now at Miami University, Oxford OH 45056, USA}
\altaffiltext{7} {now at University of Washington, Seattle WA 98195-1580}
\altaffiltext{8} {The Aerospace Corporation, Los Angeles, CA 90009, USA}
\altaffiltext{9} {Brafford \& Phillips, Batavia, OH 45103, USA} 
\altaffiltext{10} {Department of Earth and Space Science, Graduate School of Science, Osaka University, 1-1,
Machikaneyama, Toyonaka, Osaka 560-0043, Japan}
\altaffiltext{11} {Department of Astronomy, University of Washington, Box 351580 Seattle, Washington, USA}
\altaffiltext{12} {Tokyo Institute of Technology, 2-12-1 Ookayama, Meguro, Tokyo 152-8551, Japan}
\altaffiltext{13} {Division of Liberal Arts, Kogakuin University, 1-24-2, Nishi-Shinjuku, Shinjuku-ku, Tokyo, 163-8677, Japan}
\altaffiltext{14} {Astronomical Institute 'Anton Pannekoek' Science Park 904 1098 XH Amsterdam, The Netherlands}
\altaffiltext{15} {Laboratoire Lagrange, UMR7293, Universit?Le de Nice-Sophia Antipolis, CNRS, Observatoire de la C\^ote d?fAzur, 06300 Nice, France}
\altaffiltext{16} {Max Planck Institute for Astronomy, Heidelberg, Germany" with "69117 Heidelberg, K\"onigstuhl 17, Germany}
\altaffiltext{17} {Department of Physics and Astronomy, College of Charleston, 58 Coming St., Charleston, SC 29424, USA}
\altaffiltext{18} {Subaru Telescope, 650 North A'hoku Place, Hilo, HI 96720, USA}
\altaffiltext{19} {Department of Astronomy, The University of Tokyo, Hongo 7-3-1, Bunkyo-ku, Tokyo 113-0033, Japan}
\altaffiltext{20} {Institute for Astronomy, University of Hawaii, 640 North A'hoku Place, Hilo, HI 96720, USA}
\altaffiltext{21} {Department of Astrophysical Sciences, Princeton University, NJ08544, USA}
\altaffiltext{22} {Department of Astronomy, Kyoto University, Kitashirakawa-Oiwake-cho, Sakyo-ku, Kyoto, 606-8502, Japan}
\altaffiltext{23} {ExoPlanets and Stellar Astrophysics Laboratory, Code 667, Goddard Space Flight Center, Greenbelt, MD 20771 USA}
\altaffiltext{24} {Departamento de Astrof??sica, CAB (INTA-CSIC), Instituto Nacional de T?ecnica Aeroespacial, Torrej?on de Ardoz, 28850, Madrid, Spain}
\altaffiltext{25} {Institute of Astronomy and Astrophysics, Academia Sinica, P.O. Box 23-141, Taipei 106, Taiwan}
\altaffiltext{26} {Astronomical Institute "Anton Pannekoek," University of Amsterdam, Science Park 904, 1098 XH Amsterdam, The Netherlands}
\altaffiltext{27} {Institute for the Physics and Mathematics of the Universe, The University of Tokyo, Kashiwa 227-8568, Japan}
\altaffiltext{28} {Department of Cosmosciences, Hokkaido University, Sapporo 060-0810, Japan}
\altaffiltext{29} {Astronomical Institute, Tohoku University, Aoba, Sendai 980-8578, Japan}
\altaffiltext{30} {Department of Earth and Planetary Science, The University of Tokyo, Hongo 7-3-1, Bunkyo-ku, Tokyo 113-0033, Japan}
\altaffiltext{31}{Universit\"ats-Sternwarte M\"unchen, Ludwig-Maximilians-Universit\"at, Scheinerstr.~1, 81679 M\"unchen, Germany}
\altaffiltext{32}{The Graduate University for Advanced Studies , Shonan International Village Hayama-cho, Miura-gun, Kanagawa, Japan}


\begin{abstract}
One of the key predictions of modeling from the IR excess of Herbig Ae stars is that for protoplanetary disks, where significant grain growth and settling has occurred,  the dust disk has flattened to the point that it can be partially or largely shadowed by the innermost material at or near the dust sublimation radius.  
When the self-shadowing has already started, the outer disk is expected to be detected in scattered light only in the exceptional cases that the scale height of the dust disk at the sublimation radius is smaller than usual.
High-contrast imaging combined with the IR spectral energy distribution allow us to measure the degree of flattening of the disk, as well as to determine the properties of the outer disk. 
We present polarimetric differential imaging in $H$ band obtained with Subaru/HiCIAO  of one such system, MWC 480.
The HiCIAO data were obtained  at a historic minimum of the NIR excess. 
The disk is detected in scattered light from 0\farcs2-1\farcs0 (27.4-137AU).  
Together with the marginal detection of the disk from 1998 February 24 by HST/NICMOS, our data constrain the opening half angle for the disk to lie  between 1.3$\leq\theta\leq$2.2$^\circ$.  
When compared with  similar measures in CO for the gas disk from the literature, the dust disk subtends only $\sim$30\% of the gas disk scale height (H/R$\sim$0.03). 
Such a dust disk is a factor of 5-7 flatter than transitional disks, which have structural signatures that giant planets have formed. 
\end{abstract}





\section{Introduction}

For the vast majority of protoplanetary disks, modeling of the IR spectral energy distribution has been the principal tool for characterizing the properties of the dust disk. 
Models for disks derived from such data are non-unique since the viewing geometry (inclination), assumed grain opacity model, and the presence or absence of structure in the disk can combine to produce similar SEDs. 
Moreover, disks are not static: ISO and Spitzer studies of T Tauri stars have shown that variability in the 3-4.5 micron range is ubiquitous \citep{rebull11, kospal12}. 
At least some higher mass analogs to T Tauri stars, the Herbig Ae stars, also exhibit striking variability in the IR SED which has been interpreted as reflecting changes in the  scale height of the dust disk at or  near the sublimation radius \citep{sitko08}. 

For protoplanetary disks where there has been significant dust grain growth and settling, the dust disk may have flattened sufficiently to be partially or largely shadowed \citep{meeus01, dulldom04a, dulldom04b}  by material at or near the dust sublimation radius.
Under these circumstances, the amount of light reaching the outer disk is expected to be anti-correlated with the IR excess in the 1-10$\mu$m range, originating from material at or near the dust sublimation radius \citep{acke09,carciofi04,robitaille07}.  
HST and ground-based NIR single-epoch studies of Herbig Ae disks associated with IR SEDs interpreted as arising from shadowed disks (Meeus group II) resulted in a high non-detection rate, even for mm-bright, large protoplanetary disks \citep{grady05, fukagawa10}, but also in some surprising detections \citep{grady00}. 
Multi-epoch imaging of one of these stars, HD 163296, demonstrated that the illuminated portion of the disk varied with time \citep{wisniewski08}. 
These observations typically lacked simultaneous measurements of the IR SED, which would be needed to test whether the disk illumination is indeed anti-correlated with the near and mid-IR light. 

Another Herbig Ae star with documented near and mid-IR excess variability is MWC 480 (HD 31648), located at d=137$^{+31}_{-21}$ pc \citep{vanLeeuwen07}. 
This star is one of the brightest Herbig Ae stars at millimeter wavelengths \citep{mannings97}, and has been the subject of detailed studies from the X-ray through the radio regime \citep{grady10,simon00,pietu06,pietu07}, as well as repeated high-contrast imaging with HST \citep{augereau01,grady05,grady10} and NIR imaging with Subaru \citep{fukagawa10}. 
HST imaging at the epoch of ISO SWS observations resulted in a marginal scattered-light detection of this disk \citep{grady10} at a time when the IR SED suggested that the IR excess was weaker than average. 
However, the disk detection rate in scattered light is known to increase as further steps to increase the disk-to-star contrast, such as use of polarimetric differential imaging, are taken, and as the inner working angle of the observations decreases. 
We present the first firm scattered light detection of the disk of MWC 480 with an inner working angle of 0\farcs2 ~(27 AU).

\section{Observations and Data reduction}

\subsection{Subaru/HiCIAO}
We carried out H-band ($1.6 \mu $m) linear polarimetric observations of MWC 480 using the high-contrast imaging instrument HiCIAO \citep{tamura06} with a dual-beam polarimeter at the Subaru 8.2m Telescope on 2010 January 24. 
These observations were part of the Strategic Explorations of Exoplanets and Disks with Subaru \citep[SEEDS;][]{tamura09}.  
The  polarimetric observation mode acquires \textit{o}-ray and \textit{e}-ray simultaneously, and images a  field of view of  $10'' \times 20''$ with a pixel scale of 9.53 mas/pixel.  
MWC 480 was observed using a circular occulting mask with $0.3"$ diameter. 
The exposures were performed at four position angles (P.A.s) of the half-wave plate, with a sequence of P.A. = 0$^\circ$, 45$^\circ$, 22.5$^\circ$, and 67.5$^\circ$ to measure the Stokes parameters.
The integration time per wave plate position was 19.5 sec and we obtained twenty-four sets by repeating the cycle of the wave plate.
The adaptive optics system \citep[AO188; ][]{hayano08} provided a diffraction limited and mostly stable stellar PSF with FWHM of 0.07$''$ in the \textit{H} band. 
Low quality images were removed prior to production of the final images, resulting in 18 good data sets with a total integration time for
the polarization intensity (hereafter \textit{PI}) image of 351.0 sec. 
Note that the data sets are taken with ADI (Angular Differential Imaging) mode, and the field rotation is 32 degrees.

\begin{figure}
  \begin{center}
    \plotone{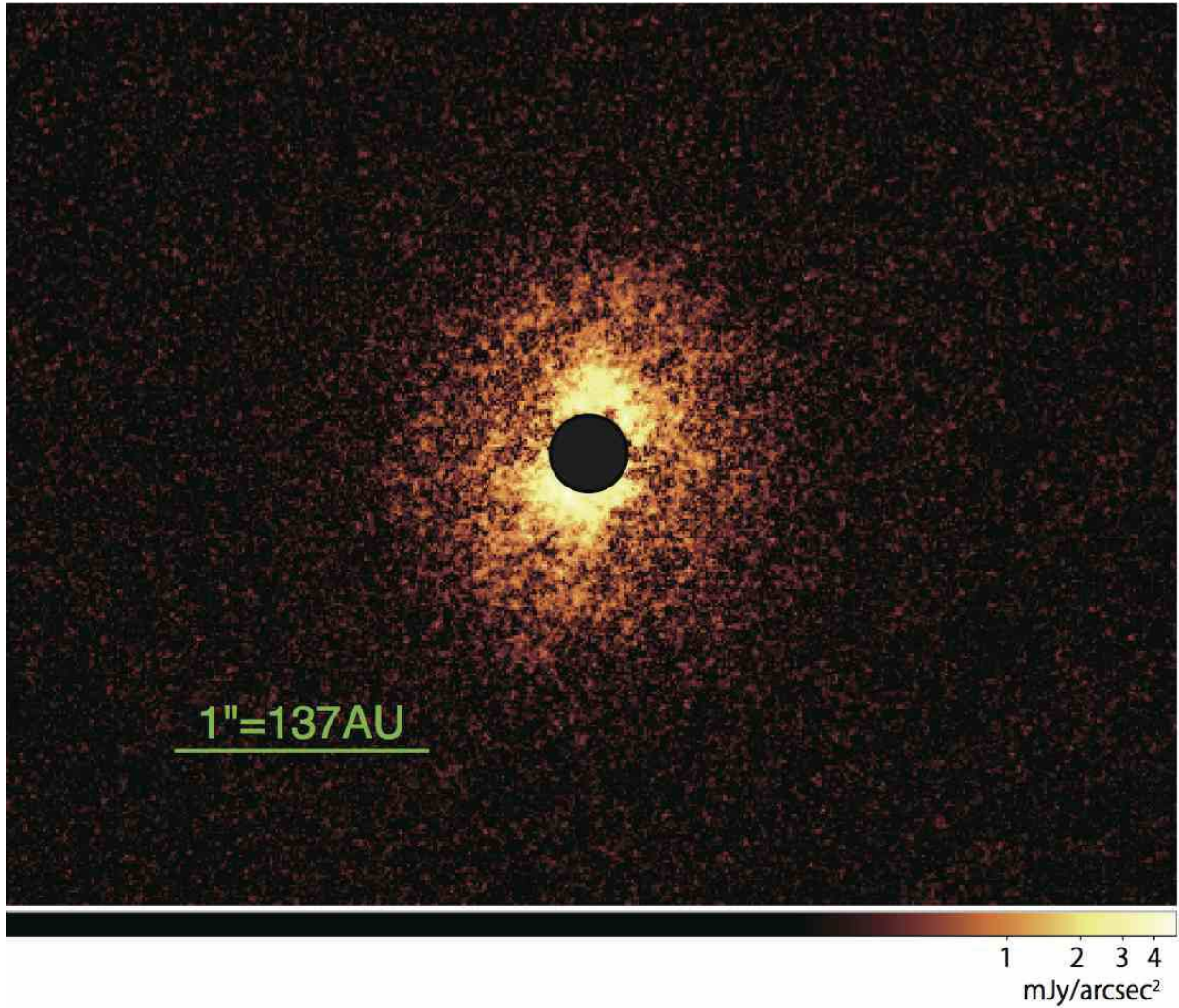}
  \end{center}
  \caption{$H$-band $PI$ image of MWC 480. Central black circle shown the occulting mask (r=0\farcs15). North is up and east to the left in this image.}
  \label{mwc480PI}
\end{figure}

The polarimetric data reduction is described in \citet{hashimoto11} and \citet{muto12}, using the standard approach for polarimetric differential imaging \citep{hinkley09}.
Instrumental polarization of HiCIAO at the Nasmyth instrument was corrected by following \citep{joos08}.
We calculated $PI$ from $\sqrt{Q^2+U^2}$. 
This approach mitigates two of the limitations of  high-contrast imagery:  the PSF observations are obtained simultaneously with the disk data,  eliminating the effects of variable seeing, and also provides a PSF measurement which is a perfect color match to the star. 
This is particularly important for systems where the J-H or H-K colors are redder than any Main Sequence candidates.  
While a PSF reference target was observed immediately after MWC 480, the seeing was gradually changing and the reference target data did not provide a good match to the PSF of the MWC 480 data. 
As a consequence, we focus on the $PI$ image rather than PSF subtraction intensity ($I$) image. 
The $PI$ image is shown in figure \ref{mwc480PI}.

Non-saturated images of MWC 480 in H band were obtained immediately after the coronagraphic sequence, using an ND filter which transmits 10\% of the light.  
Similar observations of the photometric calibration star HD 18881 (A0V, J-H=0.0$^m$=0.055, H-K=0$^m$) were obtained on 2010 January 23. 
HD 18881 was selected from the UKIRT bright standard catalog \citep{hawarden01} for the Mauna Kea filter set described by \citet{simons02}, \citet{tokunaga02}, and \citet{tokunaga05}.
We derived the zero point magnitude as 24.8 $\pm$ 0.01 from the photometry of the calibration star.

\subsection{BASS Spectrophotometry}

We observed MWC 480 during six epochs between 1996 and 2010, using The Aerospace Corporation's Broad-band Array Spectrograph System (BASS) on NASA's Infrared Telescope Facility (IRTF).  
BASS uses a cold beamsplitter to separate the light into two separate wavelength regimes. 
The short-wavelength beam includes light from 2.9-6 $\mu$m, while the long-wavelength beam covers 6-13.5 $\mu$m. 
Each beam is dispersed onto a 58-element Blocked Impurity Band (BIB) linear array, thus allowing for simultaneous coverage of the spectrum from 2.9-13.5 $\mu$m. 
The spectral resolution $R = \lambda$/$\Delta\lambda$ is wavelength-dependent, ranging from about 30 to 125 over each of the two wavelength regions \citep{hackwell90}. 
The entrance aperture of BASS is a 1-mm circular hole, whose effective projected diameter on the sky was 3\farcs3-3\farcs4, while a re-design of the dewar optics increased this value to 4\farcs0 ~in 2010. 
The observations are calibrated against spectral standard stars located close to the same airmass. 
Due to its proximity in the sky, $\alpha$ Tau usually serves as the flux calibration star.

\subsection{SpeX Cross-Dispersed Observations}

We also observed MWC 480 with the SpeX spectrograph on IRTF on 11 nights between 2006 and 2011, extending the study of \citet{sitko08}.  
The SpeX observations were made using the cross-dispersed (hereafter XD) echelle gratings in both short-wavelength mode (SXD) covering 0.8-2.4 \micron{} and long-wavelength mode (LXD) covering 2.3-5.4 \micron{} \citep{rayner09}.
These observations were obtained using a 0\farcs8 ~wide slit, corrected for telluric extinction, and flux calibrated against a variety of A0V calibration stars observed at airmasses close to that of MWC 480.  
The data were reduced using the Spextool software \citep{vacca03,cushing04} running under IDL. 

Due to the light loss introduced by the 0\farcs8 ~slit used to obtain the SXD and LXD spectra, changes in telescope tracking and seeing between the observations of MWC 480 and a calibration star may result in merged SXD or LXD spectra with a net zero-point shift compared to their true absolute flux values. 
We used a variety of techniques to check for any systematic zero-point shift in the absolute flux scale, as discussed below in greater detail. 
These included using the low dispersion prism in SpeX with a 3\farcs0 ~wide slit, JHK photometry with the SpeX guider, and the 3-13.5 $\mu$m spectrophotometry from BASS.

\subsection{SpeX Prism Observations}

On many nights we observed MWC 480 and its A0V calibration star using the low-dispersion Prism and a 3\farcs0 ~wide slit. 
To avoid saturation of the detector and minimize the wavelength calibration arc line blends, the flat-field and wavelength calibration exposures required a narrower slit, and the 0\farcs8 ~slit was used. 
On nights of good transparency and seeing 1\farcs0 ~or better, this generally produces the same flux calibration as the other methods. 
On such nights, very little light hit the slit jaws, and individual exposures showed minimal scatter. 
However, during nights of exceptionally good seeing (0\farcs7 ~or better) motion of the star on the detector can introduce small wavelength shifts, which will result in under-corrected or over-corrected telluric bands with a characteristic ``P Cygni'' shape. 
These have minimal affect on the overall absolute flux calibration, however.

\subsection{Near-IR Photometry}

On the same night that some of the spectra of MWC 480 were obtained with SpeX, images in the J, H, and K filters of the SpeX guider camera (``Guidedog'') were obtained. 
This uses the Mauna Kea filter set.

Three calibration stars were also observed. 
Two of these, HD289907 and SA-97-249, are standard stars for the Mauna Kea filter set, and the magnitudes of \citet{leggett06} were used. 
The third was HD 31069, the A0V star used for the some of the spectroscopy, where 2MASS \citep{skrutskie06,cutri03}  values were used. 
Even with short (0.12 sec) integration times per image, slight saturation effects were evident in some of the K-band data, and these were not used for the absolute flux calibration.

\section{Results}

\subsection{Disk polarization}
The disk of MWC 480 is detected  out to a radius of $\sim$1\arcsec ~(137 AU) (figure \ref{mwc480PI}).  
The central black circle indicates the 0\farcs15 radius of the occulting mask.
We measure an integrated $PI$  from 0\farcs2-2\farcs0~ of 1.2 $\pm$ 0.1 mJy, with 0.6$\pm$0.1 mJy exterior to 0\farcs6 ~in radius.  
On the same night, the star was at $H=6.74 \pm 0.08$ magnitude, resulting in a  polarized fractional intensity of 0.08\%.  
In contrast to the NICMOS marginal detection \citep{grady10}, the disk is detected at all position angles, except for a dark lane along PA$\sim$60$\pm$10$^\circ$. 
This PA is in reasonable agreement with the average disk semi-minor axis position angle of 57$\pm$2.2$^\circ$ derived from millimeter interferometry for an average of 5 CO and HCO$^+$ transitions \citep{pietu07} and the PA of the jet \citep{grady10}. 
We therefore interpret the dark band as a feature due to  depolarization  due to the grain scattering phase-function, with no indication of material deficits in the disk. 

\begin{figure}
  \begin{center}
    \plotone{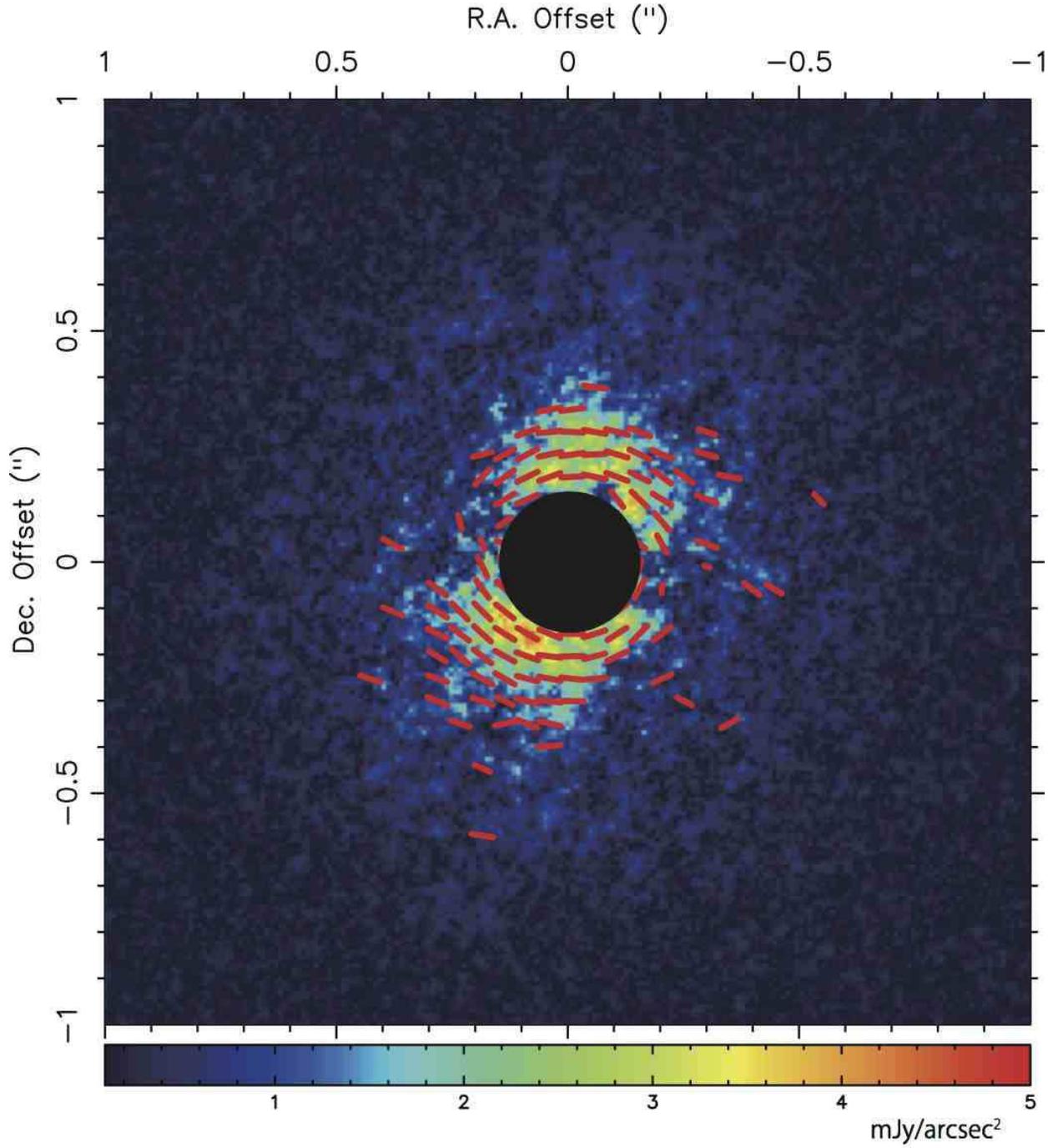}
  \end{center}
  \caption{Polarization vector map superposed on the $H$-band $PI$ image of MWC 480. Central black circle shown the occulting mask (r=0\farcs15). The vectors show just polarization angle, not polarization degrees.}
  \label{mwc480PIvec}
\end{figure}

The vector map of the $PI$ disk is shown in figure \ref{mwc480PIvec}, superposed on the bright $PI$ disk. 
The vectors show only polarization angle information, since we have no contemporary $I$ data. 
The bright inner disk ($0.2 \leq r \leq 0.4$) polarization vectors show centro-symmetric features (e.g. a butterfly pattern), indicating that the $PI$ disk is really scattered light from the central star.

\subsection{Polarization Intensity  Radial Profile}

\begin{figure}
  \begin{center}
    \plotone{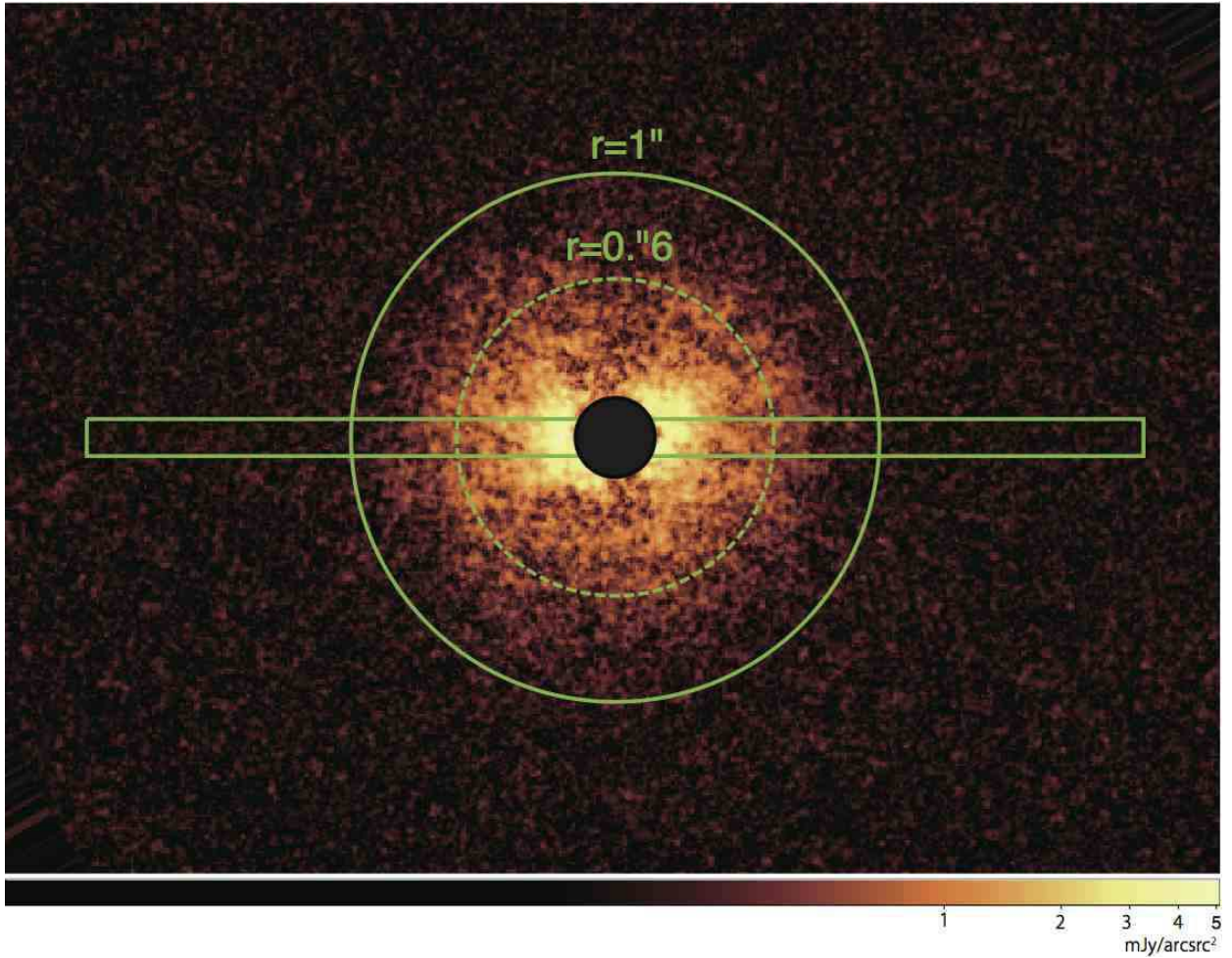}
  \end{center}
  \caption{The rotated image of polarization intensity radial profile of MWC 480.
  The vertical and horizontal PAs are 60 degrees and  150 degrees. 
  The green box is the position of radial profile, swath 14 pixel width.
  Black circle shown the occulting mask (r=0\farcs15).}
  \label{mwc480PIswath}
\end{figure}

\begin{figure}
  \begin{center}
    \plotone{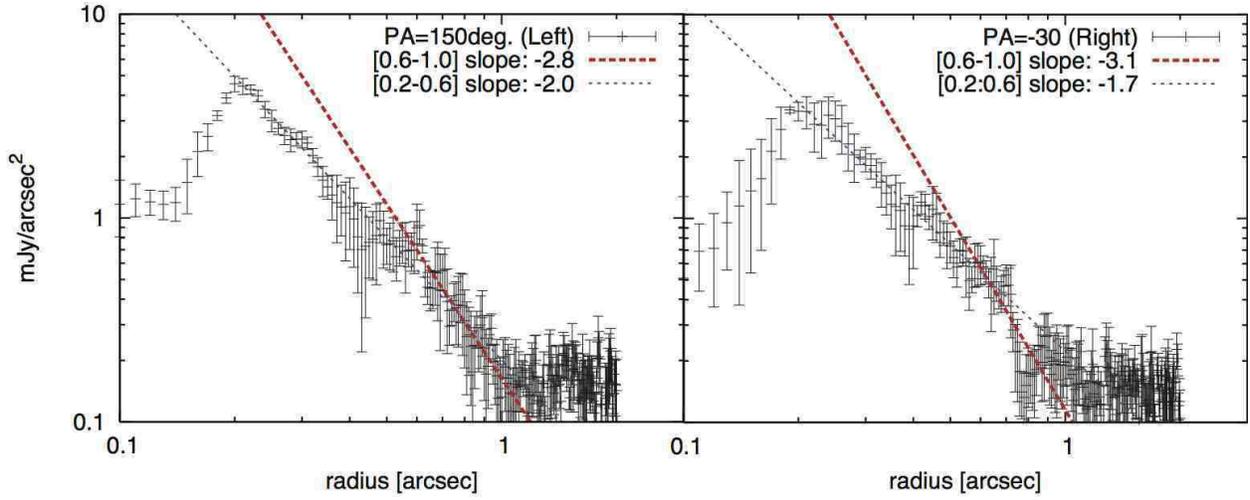}
  \end{center}
  \caption{Polarization intensity radial profile along the major axis of MWC 480.
Blue dashed lines show the profile from 0\farcs2-0\farcs6 ~of PA=150 degrees ($r^{-2.0 \pm 0.1}$ : Left) and PA=-30 degrees  ($r^{-1.7 \pm 0.1}$ : Right). 
Red dashed lines show the profile from 0\farcs6-1\farcs0 ~of PA=150 degrees ($r^{-2.8 \pm 0.2}$ : Left) and PA=-30 degrees ($r^{-3.1 \pm 0.3}$ : Right). }
  \label{mwc480PIradpl}
\end{figure}

In common with other SEEDS studies of young stellar object disks, we measure the disk radial polarized surface brightness profile along the disk major axis, or the axis of maximum elongation in the HiCIAO data. 
We extracted a swath 0\farcs13 ~wide (2x the PSF FWHM)  along the major axis from the center of the coronagraphic spot out to 3\arcsec ~on both sides of the star. 
$PI$ brightness was measured in 14 pixel $\times $ 1 pixel, 0\farcs13 ~$\times $ 0\farcs01 (figure \ref{mwc480PIswath}). 
Both radial profiles are shown in figure \ref{mwc480PIradpl}.
In the profile, r = 0\farcs15 ~is the occulting mask edge.

From 0\farcs2-0\farcs6 ~(27-84 AU) we find that the $PI$ radial surface brightness of PA=150 degrees and 30 degrees drops as $r^{-1.7 \pm 0.1}$  and $r^{-2.0 \pm 0.1}$, respectively. 
From 0\farcs6-1\farcs0 ~(84-137 AU), the radial surface brightness (SB) profile is steeper, $r^{-3.0 \pm 0.3}$. 
The outer disk radial SB profile is similar to that of HD 169142 from 123-200 AU,  \citep{fukagawa10, grady07}, HD 100546 from 10-140 AU \citep{quanz11}, and SAO 206462 \citep{muto12}. 
The shallow radial SB profile from 0\farcs2-0\farcs6 ~is similar to that seen in SAO 206462 in the region of the spiral arms where the polarization fractional intensity is 0.6 \% \citep{muto12}, although no resolved spiral features are detected in the disk of MWC 480 at the resolution of our data.

\subsection {Placing the HiCIAO detection in Context} 

\begin{figure}
  \begin{center}
    \plotone{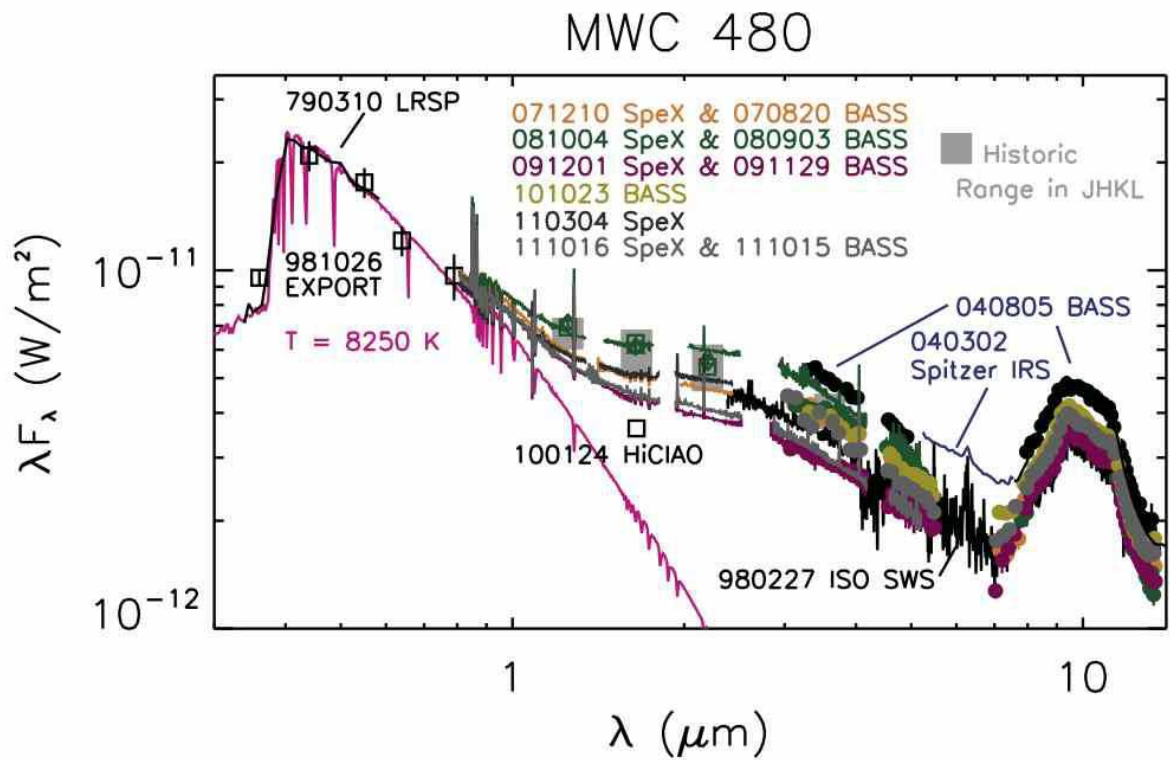}
  \end{center}
  \caption{The spectral energy distribution of MWC 480. Multi-epoch data from BASS and SpeX, along with the 2010 HiCIAO photometry are shown. Also shown are UBVRI photometry from the EXPORT consortium (Oudmeijer et al. 2001), data from the Low Resolution Spectrophotometer (LRSP) of the Pine Bluff Observatory (Sitko 1981),  spectra from the Infrared Space Observatory (ISO) and the Spitzer Space Telescope, and a T= 8250 K model atmosphere (Kurucz 1979).
Note that the HiCIAO photometry data point is at a historic minimum as compared to the ensemble of other observations presented.}
  \label{mwc480photospheric}
\end{figure}

MWC 480 has a history of correlated near- and mid-IR variability \citep{sitko08} with changes in the amplitude of the IR excess, rather than changes in stellar temperature, which has continued into the period covered by this paper.  
Individual measurements obtained from 2006 to 2011 are shown, together with the envelope of NIR variations from the literature (shaded) in figure \ref{mwc480photospheric}.  
The HiCIAO data were obtained at $H=6.74 \pm 0.08$, a historic minimum of NIR brightness for MWC 480.  
The constant temperature of the IR excess permits us to estimate the dust disk  scale  from the NIR excess. 
At the time of the HiCIAO observation, the disk scale height at the sublimation radius was only $\sim$60\% of that seen in the 1998 data, which yielded a marginal detection of the disk. 
By contrast, {\it Spitzer} IRS and BASS observations made in 2004 corresponded to a maximum excess state.  
The HiCIAO data support an anti-correlation between disk detectability and the NIR excess, consistent with predictions for dust disks where grains have grown and settled to the disk midplane \citep{sitko08, wisniewski08}. 

\section {Discussion}

Despite being the brightest Herbig Ae star at millimeter wavelengths {\citep{mannings97}}, the disk of MWC 480 has proven elusive to imaging in scattered light. 
Despite multiple attempts with HST \citep{augereau01,grady10} and Subaru/CIAO \citep{fukagawa10}, the disk has previously been imaged in scattered light only once, and then with only a marginal detection and incomplete recovery of the disk at all position angles \citep{grady10}.  
The HiCIAO $PI$ image of this disk represents the first detection of the disk in polarized light, and a detection which moreover probes the disk at all position angles, except along the system semi-minor axis.  
The successful detection of the disk relies on two factors. 
First, the HiCIAO observation was made at a time when the NIR excess was at a historic minimum, providing the most favorable illumination conditions for the outer disk. 
The second factor is  polarimetric differential imaging with simultaneous measurement of the stellar PSF from the unpolarized stellar signal. 
The $PI$ image was made by combining the Stokes $Q$ and $U$ images. 
The Stokes $Q$ and $U$ parameters were created by subtracting two split images on the HiCIAO image, which can subtract unpolarized light (PSF) of MWC 480 itself.
This approach provides not only a perfect seeing match, but also an exact color match to the star.  

Nevertheless, the disk is faint, with a fractional polarized intensity of only $\sim $0.1\%. 
Despite its relatively old age \citep[7 Myr;][]{simon00}, the disk has yet to develop  distinct  structural features, including divots, rings, or a cavity.  
Our data do reveal what appears to be a break in the radial polarized intensity power law at 0\farcs6 ~(84 AU), with a shallower power law interior to that point. 
A similarly shallow power law is seen in the region of the spiral arms for SAO 206462 \citep{muto12}.  
However, the absence of distinct structure in the disk of MWC 480 constrains any spiral features in the the disk to have  relative scale height (H/R) $<$0.05, in order for them not to be detectable at the spatial resolution of the HiCIAO data.     
In tandem with the presence of  on-going accretion, which manifests itself through a soft X-ray spectrum, FUV excess light and mass loss via  a jet \citep{grady10}, the disk of MWC 480 has a distinctly primordial character. 

As befits a dynamically cold disk, MWC 480's dust disk is also geometrically quite thin. 
Our detection of the disk, in tandem with the previous NICMOS marginal detection, allows us to better constrain where the optically thick surface of the disk lies in H band relative to the gas disk.  
\citet{pietu07} found from modeling of CO interferometric data a 10 $\pm$ 1.1 AU gas disk scale height at 100 AU, corresponding to a disk opening half angle of 6$^\circ$ and a disk inclination of 37.5$^\circ$.
\citet{grady10} estimated that the angle subtended by the dust disk rim at the sublimation radius was 2.2$^\circ$ for the 1998 NICMOS marginal detection of the disk in scattered light. 
For the HiCIAO data, the star was 0.5 magnitudes fainter at H.
If we assume that variation in the NIR excess light scales with changes in the scale height of the dust disk inner rim, the rim subtended only $\sim$60\% of the scale height seen in 1998 data.
This constrains the optically thick dust disk surface to lie above 1.3$^\circ$.
In tandem with the disk scattered light detection, this constrains the optically thick dust disk surface to lie above of 1.3$^\circ$. 
The available data therefore suggest that the surface of the dust disk lies between 1.3$\leq\theta\leq$2.2$^\circ$.
Adopting the average of this range, the dust disk has a scale height at 100 AU  only $\sim$30\% as high as the gas disk (H/R$\sim$0.03), demonstrating that this is a stratified disk.  
The synoptic data for MWC 480 further suggest that the outer dust disk is only rarely illuminated, in contrast to HD 163296 \citep{wisniewski08, fukagawa10}, although quantifying the duty cycle for disk illumination will require a denser grid of photometric and spectroscopic data than are currently available. 

\begin{figure}
  \begin{center}
    \plotone{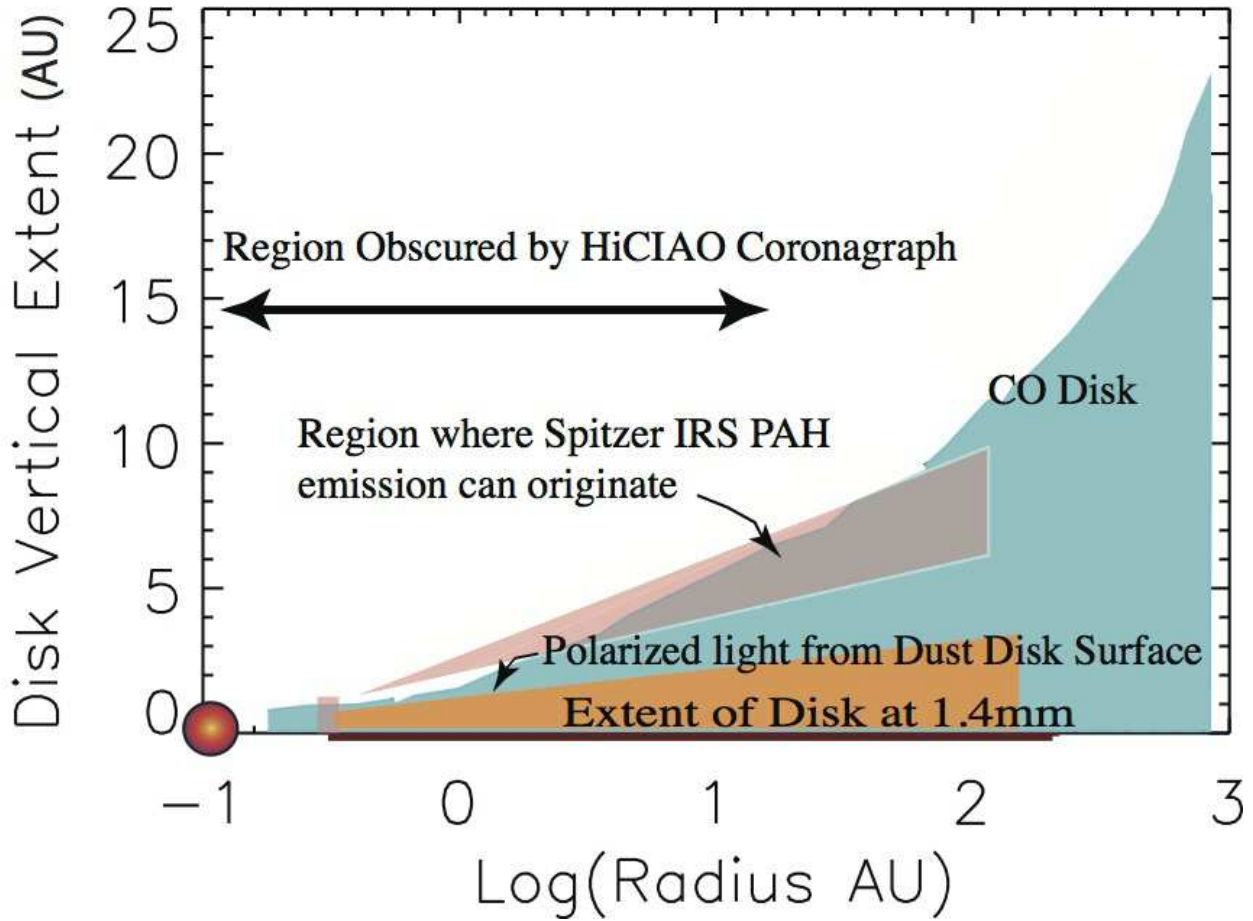}
   \end{center}
   \caption {A cartoon of the disk of MWC 480, based on data from the literature, and the geometrical constraints provided by the synoptic data and high-contrast imagery.
(brown = extent of disk at 1.4 mm; orange = dust disk surface; light blue = CO disk; pink = possible Spitzer IRS PAH emission zones.)} 
   \label{mwc480_cartoon}
 \end{figure}

A further advantage of the availability of synoptic data in tandem with more specialized observations is that we can start to extract the vertical structure of the disk. 
By combining the imagery  \citep[NICMOS 1.1$\mu$m non-detection data;][]{grady10} with the synoptic data, we find that the 2004 Spitzer IRS detection of PAH emission in MWC 480 \citep{keller08,  acke10} corresponded to a high excess state (BASS data), a factor of 3 above the ISO SWS data (figure \ref{mwc480photospheric}).
In this configuration,  the entire dust disk surface would have been shadowed, indicating that PAH emission either originates from the inner rim of the dust disk, or that it must be associated with material in the gas disk at altitude, rather than in close proximity to the outer dust disk (figure \ref{mwc480_cartoon}).  
If extended to other datasets, this technique can allow us to understand which disk constituents are routinely exposed to UV radiation, and which have less irradiation. 

We can compare the outer disk properties of MWC 480 with the roughly coeval SAO 206462, and  the younger, Solar analog LkCa 15, both of which have similar $PI$ measurements using HiCIAO \citep[Wisniewski, priv. comm.;][]{thalmann10, muto12}, and prior HST imagery. 
Both  MWC 480 and SAO 206462 have radial polarized intensity profiles beyond 0\farcs6 ~proportional to r$^{-3}$, if the polarization efficiency is constant, suggesting that the outer disks are not highly flared. 
Model fits to the SED and sub-millimeter visibility for SAO 206462 indicate a disk scale height at the inner edge of the outer disk of 9.2 AU at 46 AU \citep{andrews11}.  
To be illuminated at 100 AU, the dust disk in that system must lie above 20 AU, a factor of 6.7 times higher than that for  MWC 480.  
As for LkCa 15, the NICMOS data demonstrate that the disk is detected out to 1\farcs2 ~(Schneider, priv. comm.), with no evidence of shadowing of the outer disk along the system major axis.  
To meet this illumination condition, assuming a wall height of 7 AU at 50 AU in radius \citep{andrews11}, the dust disk of LkCa 15 must extend
more than 14 AU above the midplane at 100 AU in radius.  
The combination of a planet candidate in the disk of LkCa 15 (Kraus et al. 2011), and spiral arms which may also be  associated with Jovian-mass planets in SAO 206462 \citep{muto12}, suggests that the outer disks of these transitional disk systems are dynamically excited by massive planets. 
The extent to which this process is present may therefore constrain the masses of planets, even for systems where the inner cavity is inaccessible to current imaging technology. 
However, the primordial character of MWC 480's disk compared to coeval or even younger systems demonstrates that the process of giant planet formation may not be ubiquitous, or proceed at the same rate in all systems. 
Addressing how the planet hosting disks differ from those with no evidence of planet formation will  be a major task both for ALMA and for the next generation of high-contrast imagers.


\section{Summary}

We report the detection of the disk of the Herbig Ae star, MWC 480, in scattered, polarized  light from 0\farcs2-1\farcs0 ~(28-137 AU) in data taken with HiCIAO at the Subaru 8.2m telescope in early 2010. 
At the epoch of the detection, the NIR excess was at a historic low.  
Given a history of excess amplitude, but not temperature changes, this implies that the vertical extent of the inner rim of the dust disk, and by inference, the accretion rate onto the star,  was at a historic low. 
An anti-correlation between the scale height of the dust disk near the sublimation radius and the degree of illumination of the outer disk is expected only when the dust disk has settled toward the disk midplane. 
The data presented here, in tandem with earlier synoptic photometry and spectroscopy and a history of high-contrast imaging observations, provide direct confirmation that Herbig Ae stars with IR excesses characterized by power laws of their H/R are disks where dust grains have grown, settled, and where the outer disk is largely shadowed.  
Higher cadence synoptic observations are required to quantify the illumination/shadowing duty cycle for MWC 480, and to obtain sufficient data for other Meeus group II disks to establish whether MWC 480 is typical of the shadowed disks. 
The first steps in comparing the disk geometry of systems like MWC 480 to older, transitional disks have demonstrated that the transitional disks considered have not only wide gaps, but outer disks which are much more vertically extended than seen for MWC 480.  
As the SEEDS survey continues, we will have the opportunity to obtain data not only for other transitional disks, but also for additional primordial disks, and should be able to test the hypothesis that transitional disks have structural changes that extend over the entire disk reflecting the presence of giant planets within the gapped region.







\bigskip

This work is partially supported by KAKENHI 22000005 (MT), KAKENHI 23103004, 23740151 (MF), NSF AST 1008440 (CAG), NSF AST 1009314 (JPW), NASA NNX09AC73G (CAG and MLS) and the IR\&D program at The Aerospace Corporation (RWR). 



\clearpage

\begin{table}
  \caption{SpeX Observations}
  \label{tab:SpeX}
    \begin{center}
\begin{tabular}{lcccc}
\hline
UT Date & Mode & MWC 480 Airmass & Calibration Star Airmass & Calibration Star\\
\hline
2006.08.22 & SXD & 1.14 & 1.15 & HD 25152 \\ 
                      & LXD & 1.19  & 1.15 & HD 25152 \\
                      
2006.08.23 & SXD & 1.09 & 1.15 & HD 25152 \\
                      & LXD & 1.13 &  1.12 & HD 25152 \\
                      
2006.10.19 & SXD & 1.04 & 1.06 & HD 31592 \\  
                      & LXD & 1.04 & 1.07 & HD 31592 \\ 
                      
2006.11.28 & SXD & 1.02 &1.01 & HD 31592  \\  
                      & LXD & 1.02 & 1.02 & HD 31592  \\ 
                      
2007.02.25 & SXD & 1.18 & 1.06 & HD 31295 \\ 
                      & LXD & 1.24 &1.37 & HD 31295  \\ 
                      
2007.12.09 & SXD & 1.03 & 1.06 & HD 25152 \\  
2007.12.10 & LXD &1.28 & 1.11 & HD 25152 \\
                      & Prism & 1.05 & 1.06 & HD 25152 \\
                      
2008.10.04 & SXD & 1.12 & 1.21 & HD 31069 \\ 
                      & LXD & 1.09 & 1.23 &  HD 31069 \\
                      & Prism & 1.04 & 1.01 &  HD 31069 \\
                      
2009.12.01 & SXD & 1.20 & 1.16 & HD 31069 \\
                      & LXD & 1.16 & 1.19 &  HD 31069 \\
                      & Prism & 1.70 & 1.66 & HD 31069 \\
                      
2011.03.04 & SXD &1.04 & 1.11 & HD 31069 \\ 

2011.10.16 & SXD &1.02 & 1.07 & HD 25152\\
                     & LXD    & 1.12    &   1.14 &  HD 25152 \\
                     & Prism  & 1.02    &   1.05 &  HD 25152 \\

\hline
     \end{tabular}
  \end{center}
\end{table}
                  


\end{document}